\documentclass[journal,11pt]{IEEEtran}
\IEEEoverridecommandlockouts
\usepackage[utf8]{inputenc}
\usepackage[T1]{fontenc}
\usepackage{microtype}
\usepackage{lmodern} 
\usepackage{amsmath,amssymb}
\usepackage{graphicx}
\graphicspath{{./}}
\usepackage{booktabs}
\usepackage{tabularx}
\usepackage{array}
\usepackage{caption}
\usepackage{textgreek}
\makeatletter
\let\labelindent\relax
\makeatother
\usepackage{enumitem}
\usepackage{url}
\usepackage{hyperref}
\usepackage{textcomp}
\usepackage{footmisc}

\setlist[itemize]{noitemsep, topsep=4pt, left=1.5em}
\title{Sonus Health: Calibrated Heart-Murmur Detection from Smartphone-Based Veterinary Auscultation}

\author{%
\IEEEauthorblockN{%
\textbf{Aswin Jose}\IEEEauthorrefmark{1},
\textbf{Dr.\ Roeland P-J E.\ Decorte}\IEEEauthorrefmark{1},
\textbf{Laurent Locquet}\IEEEauthorrefmark{1}
} \\
\IEEEauthorblockA{\IEEEauthorrefmark{1}%
Decorte Future Industries Ltd / Sonus Health}
}
\begin{document}
\maketitle

\begin{abstract}
Heart disease is among the most common serious conditions in dogs and cats, and a heart murmur heard on auscultation is one of the earliest signs of such disease; such murmurs are often subtle and challenging to detect at early stages. General-practice veterinary examinations catch only a fraction of these murmurs, and definitive cardiac assessment typically requires either a board-certified cardiologist or an in-clinic echocardiogram, which may involve cost and scheduling constraints. We describe Sonus Health, a smartphone-based screening system that analyses an auscultation recording of approximately thirty seconds or longer --- captured by a pet owner at home or by a veterinarian or nurse in clinic --- and returns a tiered result within moments. The system was evaluated on 322 veterinary-labelled recordings under standard out-of-fold cross-validation. For recordings assigned to the high-confidence tier (30\% of cases), accuracy reaches 95.9\%, with 94.0\% sensitivity and 97.9\% specificity. Uncertain cases are prospectively routed to veterinary review rather than assigned an automated classification. Results are stable across standard and group-aware cross-validation, a held-out test split, and multiple random seeds. Beyond murmur detection, the platform also estimates heart rate and heart rate variability, and we position it as a screening, triage, and longitudinal-monitoring layer for companion-animal cardiac care.
\end{abstract}


\section{Introduction}
Heart disease in companion animals, including both dogs and cats, represents one of the most common medical conditions encountered in veterinary medicine, with heart murmurs among the earliest abnormalities detectable during physical examination~\cite{keene2019,luisfuentes2020}.
Detection can be challenging due to mild murmur intensity, environmental noise, variability in auscultation experience, and the well-described moderate inter-observer agreement reported among non-specialist clinicians~\cite{hoglund2004,beijken2020}.
Definitive cardiac assessment generally requires echocardiography performed by an experienced operator, often involving referral to a veterinary cardiologist or access to advanced imaging facilities; resources that may be limited by cost or availability~\cite{keene2019,luisfuentes2020}.

As a consequence, a proportion of companion-animal cardiac disease may remain unrecognised until later stages or may be detected only incidentally~\cite{paige2009}. This represents both a clinical and welfare challenge, particularly in cases where earlier recognition could facilitate monitoring, further investigation, or therapeutic intervention. An accessible system capable of assisting with cardiac screening and longitudinal monitoring, whether used at home or in clinic, could help support earlier recognition and veterinary assessment. Sonus Health was developed around this principle as a smartphone-based auscultation platform designed to analyse recordings captured both by pet owners outside the clinic and by veterinarians and nurses during clinical examination.

The platform consists of three principal components: a consumer smartphone used to acquire cardiac auscultation recordings, a companion mobile application guiding recording acquisition and review, and a cloud-based classification pipeline responsible for signal analysis. While murmur detection is the focus of the present evaluation, the same recording also supports estimation of heart rate and heart rate variability; these additional measures are described as platform capabilities and are not evaluated in this paper. A single home or in-clinic capture therefore yields a broader cardiac snapshot than murmur status alone and supports tracking these measures over repeated sessions. End-to-end latency from acquisition to preliminary result is typically under five seconds.
The platform is positioned as a screening, monitoring, and triage tool rather than a stand-alone diagnostic modality and was engineered around the principle that calibrated uncertainty is preferable to confidently incorrect automated classification within a clinical workflow.

\section{Data}
The evaluation reported here is based on a cohort of real-world auscultation recordings. Every recording was produced under the platform's intended operating conditions and captured with a consumer smartphone microphone. Some were recorded by pet owners outside the clinic and others by veterinarians or nurses during clinical examination, rather than re-recorded under idealised laboratory conditions. This is a deliberate choice: performance reported on cohorts re-captured under ideal conditions may overstate deployed performance, because the recording distribution at evaluation can differ systematically from the distribution at use.

Each recording carries a binary murmur label assigned by a veterinary cardiologist who was blinded to the model's predictions. No automated or machine-learning step was involved in labelling, so the labels used to evaluate the platform are independent of the platform itself.
Recordings are typically approximately thirty seconds or longer and are acoustically representative of real-world capture in both home and clinical settings, with respiratory sounds, handling noise, and ambient environmental sound frequently present; no post-hoc cleaning was applied beyond the platform's routine pre-processing.
All recordings were collected from consenting platform users under the data-handling terms agreed to at account creation and have been anonymised at the animal and owner level prior to inclusion in the evaluation cohort.

The evaluation cohort comprises 322 auscultation recordings drawn from a single April 2026 production export, of which 149 are murmur-positive and 173 are murmur-negative. In this evaluation cohort, the positive-class prevalence was approximately 46\%, close enough to balanced that accuracy and area under the ROC curve can be interpreted without prevalence-adjustment. The recordings span both dogs and cats and were captured in both home and clinical settings. They are drawn from 253 distinct animals; 216 of the 253 contribute a single recording, and a small minority contribute several across separate sessions, with the maximum count from any single animal being nine. The animal identifier is used as a grouping key so that, in the group-aware protocol below, every recording from a given animal is constrained to the same side of every split.

\section{Model}
The classification pipeline combines two probabilistic models that view each recording through materially different lenses, and the underlying design principle of the platform is that their combination improves performance compared with either model alone. The first model operates on a frequency-domain representation of the audio and is sensitive to the fine-grained spectral and temporal structure that characterises murmurs; it is strong on recordings where the murmur is acoustically present but subtle, and where the clinically relevant signal is distributed across the cardiac cycle rather than concentrated at obvious peaks. The second model operates on a summary feature vector derived from the full recording and is sensitive to session-level acoustic, timing, and signal-quality properties that the first model does not see directly; it is strong on recordings of conventional murmur morphology and is more robust to noise because the summary statistics it consumes average out transient disturbances. The two failure modes are, in practice, largely disjoint, which is the property that makes their combination valuable: a murmur that is acoustically present but summary-statistic-invisible is typically caught by the first model and missed by the second, while a murmur obscured by noise is typically caught by the second and missed by the first.

Before the two probabilistic outputs can be combined, they must be placed on a common probability scale: the first model's output is passed through a rank-preserving calibration step that places its probability scale in correspondence with the second model's. Calibration does not change the discriminative ordering of predictions; the first model's accuracy in isolation is unchanged by this step; but it standardises the numerical interpretation, so that a probability of 0.8 from the first model carries the same meaning as a probability of 0.8 from the second. Without this step, downstream thresholding on the combined output would be operating on two different scales at once, and the tiered post-processing rule described below would not behave as intended. The calibration is fit on out-of-fold predictions from the first model rather than on its training data, which keeps the calibrator from inheriting any optimism that the model may have on the recordings it was trained on.

The two calibrated probabilities are combined through a deterministic post-processing rule that assigns every recording to one of three confidence tiers. Recordings on which both models strongly agree on the same outcome are placed in a high-confidence tier and returned directly to the owner. Recordings on which the two models disagree on the outcome are placed in a low-confidence tier and routed to veterinary review without the platform committing to a call; this mechanism reduces the risk of confidently incorrect predictions, an important consideration for an automated clinical screening tool. Intermediate cases, where both models agree on the outcome but at least one is not strongly confident, are placed in a moderate-confidence tier and returned to the owner with a visible caveat recommending clinical confirmation. This document focuses on the platform's validation methodology and deployed performance; architectural details, feature specifications, training procedures, calibration mechanics, and the specific threshold values that define the tiers are proprietary to Sonus Health.

\section{Results}

\begin{figure*}[t]
\centering
\includegraphics[width=0.82\textwidth]{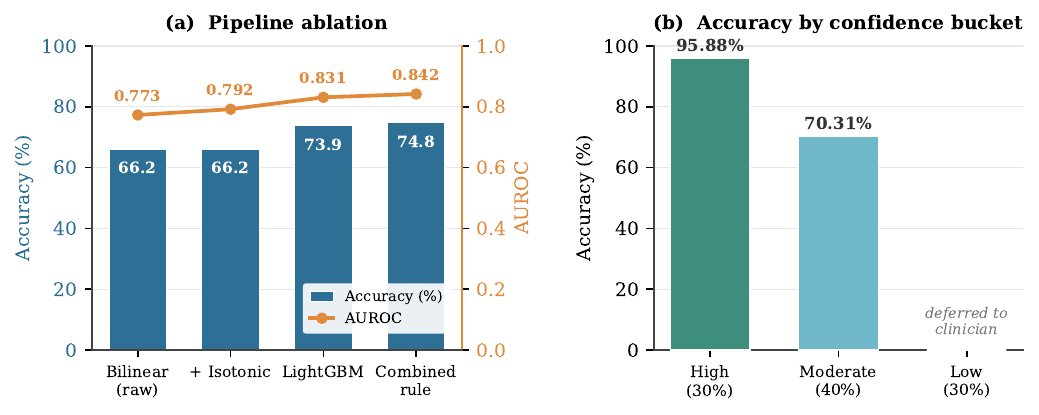}
\caption{Benchmark results on the standard five-fold pooled out-of-fold evaluation ($n=322$). (a)~Accuracy (bars, left axis) and AUROC (line, right axis) across the four pipeline stages: isotonic calibration lifts AUROC at unchanged accuracy, the tabular stream carries most of the accuracy, and the agreement rule adds precision and a final point of AUROC. (b)~Accuracy by prospective confidence bucket; the low (model-disagreement) bucket is deferred to veterinary review and is therefore not scored.}
\label{fig:results}
\end{figure*}

Performance was measured under four protocols, each designed to test a different aspect of model robustness, all seeded identically for reproducibility. The standard five-fold cross-validation (\texttt{StratifiedKFold}) gives the primary pooled out-of-fold read-out. To ensure results are not inflated by recordings from the same animal appearing on both sides of a split, the group-aware five-fold cross-validation (\texttt{StratifiedGroupKFold}, keyed on animal identifier) repeats the analysis with every recording from an animal constrained to one side of each split. A group-aware 70/15/15 train--validation--test holdout reserves a 46-recording test set that informs no hyperparameter, providing a generalisation sanity check. A seed-stability analysis reruns the group-aware protocol over seeds 42, 1337, 2026, and 7. Every metric is reported with percentile bootstrap 95\% confidence intervals from 2{,}000 resamples stratified within class, and the pipeline regenerates byte-for-byte from a fixed seed. We report a per-component ablation (Table~\ref{tab:ablation}) and a prospective confidence-bucket stratification (Table~\ref{tab:buckets}).

Under standard cross-validation the combined pipeline reaches 74.84\% accuracy (95\% CI [70.2, 79.5]) with an AUROC of 0.8415. The per-component ablation shows where this comes from. The bilinear network in isolation is recall-dominant but has lower precision (accuracy 66.15\%, $F_1$ 0.714, AUROC 0.773, precision 0.586, recall 0.913). Asymmetric isotonic calibration leaves its accuracy and $F_1$ untouched but lifts AUROC by 1.9 points to 0.792 --- representing an improvement in probability calibration while preserving classification behaviour, and the step that makes the downstream extreme-band cutoffs meaningful. The tabular LightGBM model is the stronger single stream (accuracy 73.91\%, AUROC 0.831, precision 0.692), beating the calibrated bilinear network on accuracy by 7.8 points and on precision by 10.6. The agreement rule matches the tabular model's accuracy and adds a further point of AUROC to reach 0.842, while raising precision to 0.743; it commits only when both calibrated probabilities agree and otherwise defers. Under group-aware cross-validation the same pipeline reaches 74.53\% accuracy and AUROC 0.8371 --- a difference of 0.31 and 0.44 points respectively, well within fold-to-fold noise, suggesting that animal-level leakage is not a material driver, which is unsurprising given that 216 of 253 animals contribute a single recording.

A clinically relevant breakdown is by confidence bucket. The high-confidence bucket --- assigned prospectively to recordings on which both calibrated probabilities fall in the same extreme band --- covers 30.1\% of the cohort ($n=97$) and reaches 95.88\% accuracy, with 94.0\% sensitivity, 97.9\% specificity, and 0.979 precision (47 true positives, 1 false positive, 3 false negatives, 46 true negatives). This is the operating point the product runs at on the cases it returns directly to the owner, and it approaches the performance reported for clinician review in comparable studies. The moderate-confidence bucket covers 39.8\% ($n=128$) and reaches 70.3\% accuracy, which remains useful when presented with a visible caveat. The low-confidence bucket, the remaining 30.1\% ($n=97$), is the model-disagreement set: its 0.30 sensitivity is not a failure but the explicit mechanism by which the system declines to call ambiguous recordings and routes them to veterinary review instead. Under group-aware cross-validation the high bucket is essentially unchanged (95.65\% accuracy at 28.6\% share), confirming the operating point is not an artefact of the split.

Figure~\ref{fig:results} summarises both views: the left panel traces accuracy and AUROC across the four pipeline stages, and the right panel shows accuracy by confidence bucket against its cohort share. The headline numbers are stable across protocols and seeds. Across four random seeds the combined rule varies by only $\pm$0.3 points in accuracy ($0.745 \pm 0.003$) and $\pm$0.4 points in AUROC ($0.841 \pm 0.004$); the combined stage is in fact \emph{more} stable than either single stream, because the agreement rule produces a prediction only when both streams agree, damping per-stream variance. On the held-out test split ($n=46$), which informs no hyperparameter, the combined rule reaches 84.78\% accuracy and the tabular stream reaches AUROC 0.947; the small test size gives correspondingly wide bootstrap intervals, so these are reported as a sanity check rather than the authoritative read-out, which remains the cross-validation estimate. The evaluation pipeline is reproducible: all three artifacts regenerate identically from a fixed seed, feature extraction is deterministic, and the combined artifact set totals under 4~MB.

\begin{table}[t]
\centering
\caption{Per-component ablation, standard five-fold pooled OOF ($n=322$). Calibration is rank-preserving, so it lifts AUROC without changing accuracy or $F_1$; the agreement rule matches the tabular accuracy and adds precision and AUROC.}
\label{tab:ablation}
\renewcommand{\arraystretch}{1.25}
\begin{tabularx}{\columnwidth}{@{}Xcccc@{}}
\toprule
\textbf{Stage} & \textbf{Acc.} & \textbf{$F_1$} & \textbf{AUROC} & \textbf{Prec.} \\
\midrule
Bilinear (raw) & 66.15\% & 0.714 & 0.773 & 0.586 \\
\quad + isotonic calibration & 66.15\% & 0.714 & 0.792 & 0.586 \\
LightGBM (tabular) & 73.91\% & 0.736 & 0.831 & 0.692 \\
Combined agreement rule & \textbf{74.84\%} & 0.720 & \textbf{0.842} & \textbf{0.743} \\
\bottomrule
\end{tabularx}
\end{table}

\begin{table}[t]
\centering
\caption{Prospective confidence-bucket stratification of the combined rule (standard CV, $n=322$). The low (model-disagreement) bucket is deferred to veterinary review; its low sensitivity is the deferral mechanism by design. Accuracy, sensitivity, and specificity are not reported for this bucket because these recordings are deferred rather than returned as platform predictions, so there is no platform decision to score.}
\label{tab:buckets}
\renewcommand{\arraystretch}{1.25}
\begin{tabularx}{\columnwidth}{@{}Xccccc@{}}
\toprule
\textbf{Bucket} & \textbf{$n$} & \textbf{Share} & \textbf{Acc.} & \textbf{Sens.} & \textbf{Spec.} \\
\midrule
High & 97 & 30.1\% & \textbf{95.9\%} & 0.940 & 0.979 \\
Moderate & 128 & 39.8\% & 70.3\% & 0.763 & 0.652 \\
Low (deferred) & 97 & 30.1\% & --- & --- & --- \\
\bottomrule
\end{tabularx}
\end{table}

\section{Discussion and Conclusion}

\begin{table*}[t]
\centering
\caption{Comparison to the most directly comparable systems. \emph{This is a qualitative side-by-side only}: differences in species, acquisition device, labelling protocol, and prevalence preclude a direct quantitative comparison.}
\label{tab:priorart}
\renewcommand{\arraystretch}{1.25}
\begin{tabularx}{\textwidth}{@{}Xlrllr@{}}
\toprule
\textbf{System} & \textbf{Species} & \textbf{$n$} & \textbf{Device} & \textbf{Headline metric} & \textbf{Value} \\
\midrule
PhysioNet 2022 top team~\cite{reyna2023} & Human (paediatric) & 5{,}272 & Clinical stethoscope & Weighted acc.\ / AUROC & 0.780 / $\approx$0.96 \\
McDonald 2024~\cite{mcdonald2024} & Canine & $\approx$790 & Electronic stethoscope & AUROC & 0.926 \\
Bisgin 2025~\cite{bisgin2025} & Canine & 70 & Electronic stethoscope & Sens.\ / Spec.\ (mild) & 0.910 / 0.937 \\
\textbf{This work} (combined, group CV) & \textbf{Canine + feline} & \textbf{322} & \textbf{Smartphone microphone} & \textbf{AUROC} & \textbf{0.837} \\
\textbf{This work} (high bucket) & \textbf{Canine + feline} & \textbf{322} & \textbf{Smartphone microphone} & \textbf{Accuracy} & \textbf{0.959} \\
\bottomrule
\end{tabularx}
\end{table*}

Three design decisions contribute substantially to the platform's measured behaviour, with each supported by the evaluation results. The decision to run two independent streams rather than a single model accepts additional complexity in exchange for largely disjoint failure modes: the bilinear network is recall-dominant and noisy (precision 0.586), the tabular model is precision-strong but blind to subtle spectro-temporal cues, and the agreement rule inherits the tabular precision while gaining a calibration check from the bilinear stream. The decision to calibrate the bilinear output before combination shifts its AUROC by less than two points but has a much larger product-facing effect: the agreement rule's 0.20/0.75 cutoffs are only meaningful on a calibrated scale, and without this step the high bucket's 95.88\% accuracy would be less reliable, because the two classifiers would be operating on different probability scales. The decision to return a bucketed output rather than a binary decision accepts that the platform declines to commit on roughly 30\% of recordings in exchange for a high-confidence operating point at 0.979 precision and 0.979 specificity. Each choice is conservative on its own; none maximises a single headline metric; but together they produce an operating profile suited to a screening setting.

Relative to the published companion-animal literature, the platform's overall AUROC of 0.84 trails the most directly comparable clinician-grade benchmark --- McDonald 2024's board-certified-cardiologist canine cohort at AUROC 0.926 --- by roughly nine points, a gap consistent with the shift from electronic-stethoscope capture to consumer-smartphone capture, often under real ambient noise. The aggregate comparison is informative, but a complementary view is at the level of the high-confidence bucket. On the high-confidence bucket the platform reaches 95.9\% accuracy, within noise of the clinician-cohort benchmark, while explicitly deferring the recordings it cannot confidently call. To our knowledge, few prior canine or feline studies report a prospectively-assigned confidence-bucket output of this kind; the closest analogue is Bisgin 2025's per-intensity breakdown, which is a retrospective slicing conditional on ground truth rather than a prediction-time signal. The platform's bucket assignment is, by contrast, decided at prediction time from the agreement of the two calibrated streams, which makes the high-confidence operating point a genuine deployment property rather than a retrospective slice of a correct-answer subset. This is the property that makes the platform usable as a clinical triage layer rather than only a research artefact.

The platform also addresses a different clinical need from specialist evaluation by supporting accessible screening; the practical alternative most pet owners actually face is general-practice auscultation rather than cardiologist review. Inter-rater agreement between non-specialist veterinarians on auscultation is well-documented as moderate, with detection of subtle and early-stage murmurs particularly inconsistent. The platform delivers 95.9\%-accurate, 97.9\%-specific results on the roughly 30\% of recordings it routes to the high-confidence bucket --- whether those were captured by an owner outside the clinic or by a clinician during examination --- and routes the remaining cases either to a result with a visible caveat or to direct veterinary review. The intended clinical effect is to support earlier identification of potential murmurs, provide additional information to owners, and help prioritise cases requiring veterinary attention. Because the same low-cost capture is designed to support heart-rate and heart-rate-variability estimates and can be repeated cheaply over time, the platform additionally supports longitudinal monitoring of an individual animal rather than only a one-off screen. This is not an incremental improvement on existing screening; it is a new screening tier that did not previously exist in the companion-animal cardiac workflow.

Several limitations temper these results. The evaluation cohort is modest in size (322 recordings from 253 animals, drawn from a single April 2026 production export), and although results are stable across cross-validation protocols, seeds, and a held-out split, the held-out test set is small ($n=46$) with a correspondingly wide confidence interval. The cross-validation estimates are also retrospective; prospective validation on newly collected recordings, ideally across multiple sites and devices, remains future work. The comparison to prior systems in Table~\ref{tab:priorart} is qualitative only, as species, acquisition device, labelling protocol, and prevalence differ across studies. Finally, deployment considerations --- including robustness across the full range of consumer hardware, capture technique by untrained owners, and longitudinal label drift --- are not characterised here and warrant dedicated study.

The robustness of the headline numbers is, from a credibility perspective, as important as their magnitude. The combined pipeline produces near-identical accuracy and AUROC under standard and group-aware cross-validation (74.84\%/0.842 vs 74.53\%/0.837), varies by 0.3 points or less across four random seeds, and produces a held-out point estimate that sits inside the cross-validation bootstrap interval. The benchmark regenerates byte-for-byte from a fixed seed, the calibration step is provably rank-preserving, and the high-confidence operating point is conditioned only on model agreement rather than any retrospective use of ground truth. The combination of measured, reproducible clinical performance, a bucketed output that acknowledges its own uncertainty, and a deliberate deferral channel is what supports further clinical evaluation and deployment of the platform. It is not a replacement for a cardiologist, and does not claim to be. It is a screening and triage layer that sits in front of the clinical workflow, returns a result the veterinarian can trust on the cases where it commits, and explicitly defers the rest; which is exactly the operating profile a deployable veterinary AI system should have.


\end{document}